# Anomalous electrical transport and magnetic skyrmions in Mn-tuned Co$_9$Zn$_9$Mn$_2$ single crystals

Fangyi Qi[1], Yalei Huang[1], Xinyu Yao[1], Wenlai Lu[1, *], Guixin Cao[1, 2, *]

[1]*Materials Genome Institute, Shanghai University, 200444 Shanghai, China*

[2]*Zhejiang Laboratory, Hangzhou 311100, China*

**Abstract**

*β*-Mn-type Co$_x$Zn$_y$Mn$_z$ ($x + y + z = 20$) alloys have recently attracted increasing attention as a new class of chiral magnets with skyrmions at and above room temperature. However, experimental studies on the transport properties of this material are scarce. In this work, we report the successful growth of the *β*-Mn-type Co$_{9.24}$Zn$_{9.25}$Mn$_{1.51}$ and Co$_{9.02}$Zn$_{9.18}$Mn$_{1.80}$ single crystals and a systematic study on their magnetic and transport properties. The skyrmion phase was found in a small temperature range just below the Curie temperature. The isothermal ac susceptibility and dc magnetization as a function of magnetic field confirm the existence of the skyrmion phase. A negative linear magnetoresistance over a wide temperature range from 2 K to 380 K is observed and attributed to the suppression of the magnetic ordering fluctuation under high fields. Both the magnetization and electrical resistivity are almost isotropic. The quantitative analysis of the Hall resistance suggests that the anomalous Hall effect of Co$_{9.24}$Zn$_{9.25}$Mn$_{1.51}$ and Co$_{9.02}$Zn$_{9.18}$Mn$_{1.80}$ single crystals is dominated by the intrinsic mechanism. Our findings contribute to a deeper understanding of the properties of Co$_x$Zn$_y$Mn$_z$ ($x + y + z = 20$) alloys material and advance their application in spintronic devices.

---

* Corresponding author, Email: guixincao@shu.edu.cn and wenlai@t.shu.edu.cn

## I. INTRODUCTION

Skyrmions have captured vast attention due to their various intriguing properties including the topological Hall effect (THE), the emergent electrodynamics, as well as their potential applications in low-power electronics and non-volatile memories[1-8]. Over the last decade, a large amount of research on skyrmions has been conducted on thin films or quasi-two dimensional systems[9]. Nevertheless, three-dimensional skyrmion systems have richer topological magnetic structures, whose formation and dynamics will pave the way for three-dimensional spintronic devices for brain-inspired computing[10]. Among the three-dimensional systems, chiral magnets exhibit relatively small size (<100 nm) of skyrmions, thus are more appealing for enabling functional skyrmionic memory devices with higher density[10]. However, the sub-ambient temperature of skyrmions formation in most chiral magnets limits their practical application[11-14]. The found of room temperature skyrmions breaks the application limitations imposed by the low formation temperature of skyrmions in other chiral magnets[2, 11-15]. The recent discovery of skyrmions at and above room temperature in $\beta$-Mn-type $Co_xZn_yMn_z$ ($x + y + z = 20$) chiral magnet, provides a significant step towards application[15, 16]. In $Co_8Zn_{10}Mn_2$, the electrical manipulation of skyrmions at room temperature has been realized by current pulses[17]. More interestingly, metastable skyrmions have been observed in $Co_8Zn_8Mn_4$, with a much wider temperature and field range beyond the narrow region of the equilibrium skyrmion phase, thus further expanding the limited set of their application[18].

Despite the extensive studies on $Co_xZn_yMn_z$ ($x + y + z = 20$) compound[17-21], most of the current research focuses on the magnetic configuration by small-angle neutron scattering (SANS), Lorentz transmission electron microscopy (LTEM) measurements and other sophisticated techniques that require large-scale instrument, with little attention to the electrical transport properties. However, the electrical transport and magnetic properties are crucial for the real applications of skyrmion-based spintronic devices. On the other hand, the detection of skyrmions can also be probed by magnetoresistance in spin valves or magnetic tunnel junctions, non-collinear magnetoresistance as well as Hall measurement[4, 10, 22]. Actually, anomalous Hall effect (AHE) is one of the important means and tools to explore and characterize the transport properties of ferromagnetic materials. As far as we know, however, there is only one study on the electrical transport properties of the skyrmion host material $Co_xZn_yMn_z$ [23]. In that study [23], the observed AHE of the polycrystalline $Co_7Zn_8Mn_5$ was mainly attributed to the skew scattering. However, it is known that the skew scattering dominates the AHE only in the high conductivity regime where $\sigma_{xx} \geq 10^6 (\Omega\,cm)^{-1}$[24]. Considering the reported conductivity of about $5 \times 10^3 (\Omega\,cm)^{-1}$ that falls far away out of the high conductivity regime, the claim of the dominant skew scattering for the polycrystalline $Co_7Zn_8Mn_5$ needs revisiting [23]. Therefore, in order not only to figure out the mechanisms that dominate the AHE, but also to serve as a supplement to the current research on this skyrmion host material for potential applications in spintronic, study on the electrical transport properties of the $Co_xZn_yMn_z$ ($x + y + z = 20$) compound is essential.

In this work, the high-quality $Co_9Zn_9Mn_2$ single crystals were synthesized by the self-flux method. We systematically investigated the magnetic and transport properties of the samples. The magnetic susceptibility results show that skyrmions appear in the temperature range from 354 K to 360 K. Moreover, the longitudinal magnetoresistance (LMR) and transverse magnetoresistance (TMR) both display negative magnetoresistance, which is associated with the suppressed fluctuation of magnetic ordering by magnetic field. Importantly, we find that a detailed analysis about the Hall resistance suggests a main contribution of the intrinsic mechanism to the observed AHE in $Co_9Zn_9Mn_2$. This assumption is further confirmed by the linear relationship between intrinsic anomalous Hall conductivity (AHC) and the magnetization, which is different from the previously reported skew scattering of $Co_7Zn_8Mn_5$ alloy[23]. Our works provides fundamental transport and magnetic understanding for the high-temperature skyrmions host materials.

## II. EXPERIMENTAL DETAILS

Single crystals of $Co_xZn_yMn_z$ ($x + y + z$) = 20 with different Mn contents were synthesized by using the Zn-flux method. Co (99.99%), Zn (99.99%) and Mn (99.99%) grains were mixed and placed in an alumina crucible. The crucible was then sealed in a quartz tube in a vacuum environment, and heated at 1100 °C for 2 days in the furnace before slowly cooled down to 900 °C. It was kept at 900 °C for 2 days before finally water quenched to get the samples. The resulting single crystals are very shiny with no regular shape as shown in the insets of Fig. 1(b) and (c). X-ray diffraction (XRD) was conducted by using the X-ray diffractometer (XRD, Bruker D2 PHASER) and Bruker D8 Discovery with home-made high-throughput attachment. The chemical compositions were characterized by energy-dispersive X-ray spectroscopy (EDX, HGSTFlexSEM-1000). The crystals utilized in this study have compositions of $Co_{9.24}Zn_{9.25}Mn_{1.51}$ and $Co_{9.02}Zn_{9.18}Mn_{1.80}$, respectively. Magnetization measurements of single crystals were performed in a SQUID magnetometer (MPMS, Quantum Design). The resistivity was measured by a standard four-probe method using a Physical Property Measurement System (PPMS-14, Quantum Design).

**III. RESULTS AND DISCUSSION**

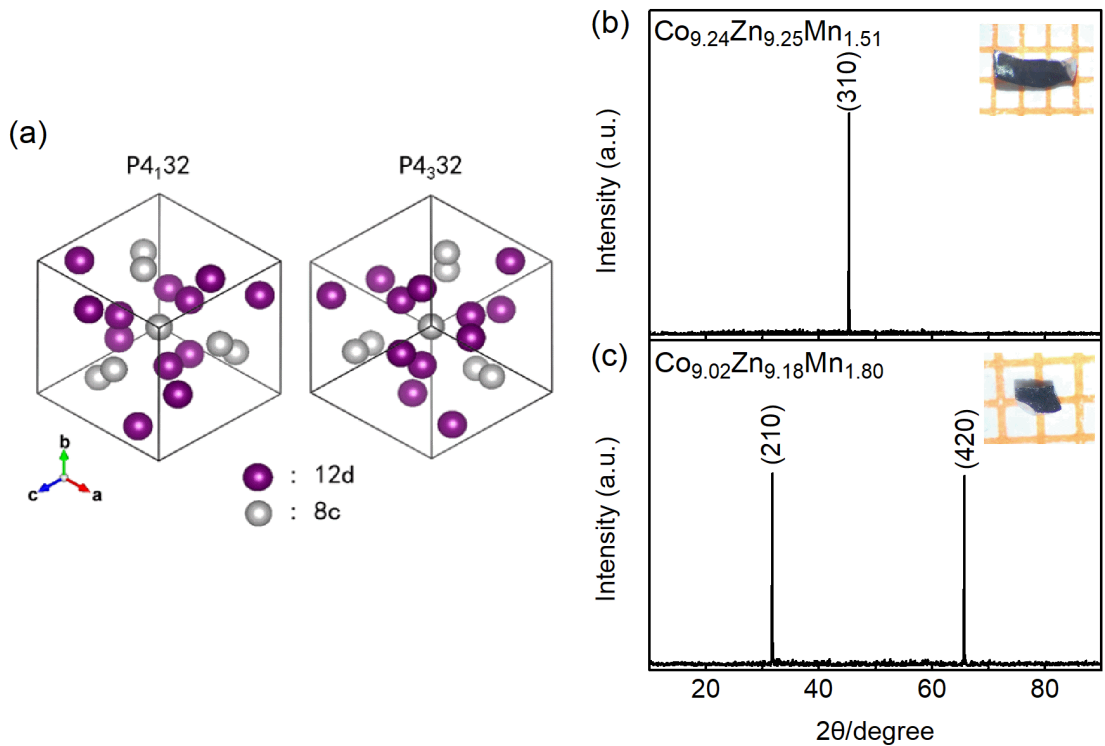

**Fig. 1.** (a) Schematic of $\beta$-Mn-type structure with $P4_132$ and $P4_332$ space group. (b)

X-ray diffraction pattern of $Co_{9.24}Zn_{9.25}Mn_{1.51}$ single crystal. (c) X-ray diffraction pattern of $Co_{9.02}Zn_{9.18}Mn_{1.80}$ single crystal. Inset: Photo of samples.

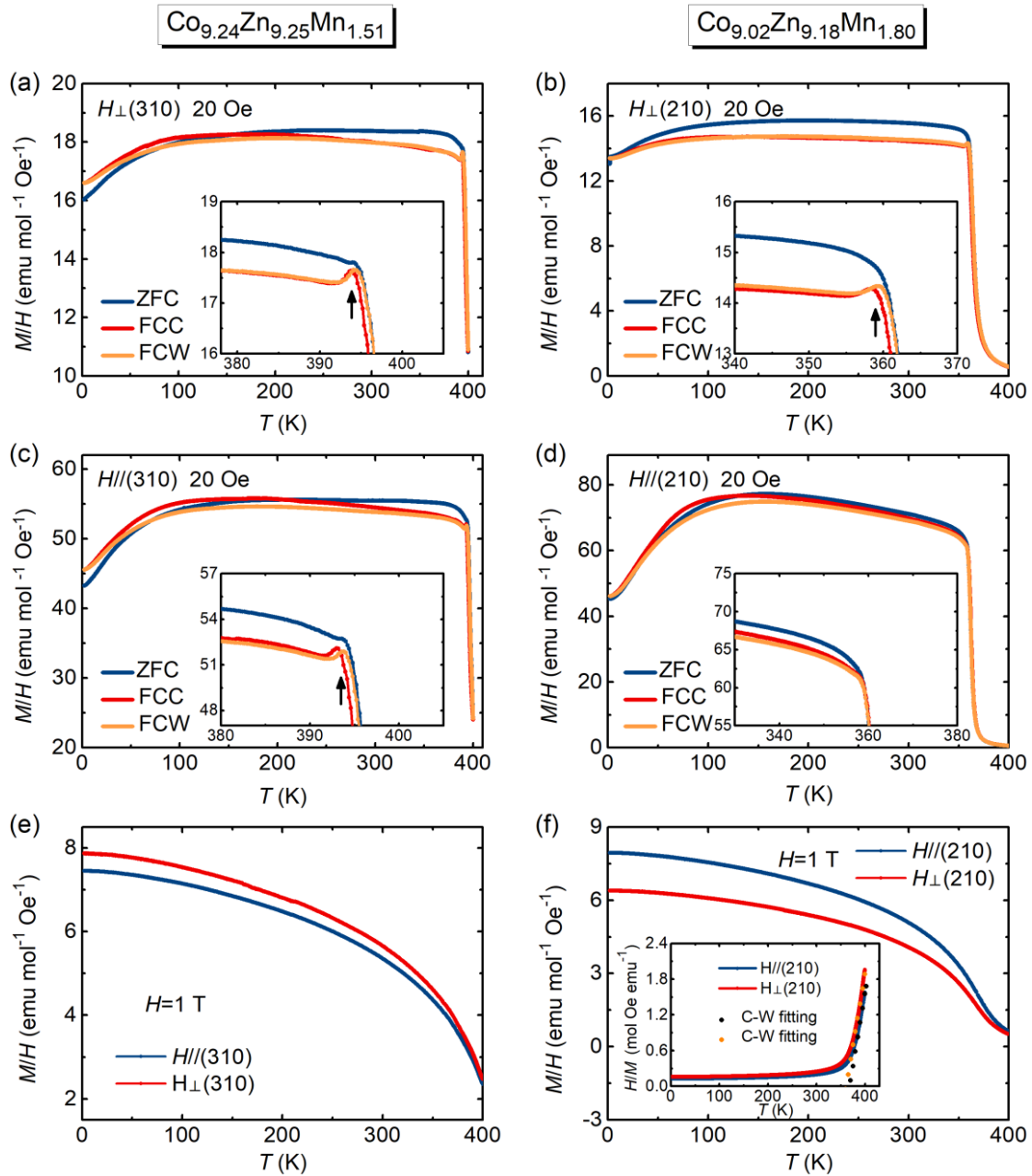

**Fig. 2.** (a)-(d) Temperature dependence of magnetic susceptibility $M/H$. Insets show the magnetic anomaly around the Curie temperature. (e) and (f) Temperature dependence of $M/H$ for samples under $\mu_0H=1$ T. Inset shows the inverse susceptibility was fitted by the Curie-Weiss formula.

XRD of powdered single crystals indicates a chiral $\beta$-Mn-type structure for both

Co$_{9.24}$Zn$_{9.25}$Mn$_{1.51}$ and Co$_{9.02}$Zn$_{9.18}$Mn$_{1.80}$, which is similar with that of parent Co$_9$Zn$_9$Mn$_2$. XRD pattern of single crystal display sharp diffraction peaks as shown in Fig.1(b) for the (310) lattice plane of Co$_{9.24}$Zn$_{9.25}$Mn$_{1.51}$ and in Fig.1(c) for the (210) plane of Co$_{9.02}$Zn$_{9.18}$Mn$_{1.80}$, respectively, confirming the high quality of our single crystals. As depicted in Fig.1(a), Co$_x$Zn$_y$Mn$_z$ ($x + y + z$) = 20 compounds crystallizes in a chiral $\beta$-Mn-type structure with space group $P4_132$ or $P4_332$ and have 20 atoms in each unit cell[15]. This structure contains two crystallographic sites, 8$c$ sites and 12$d$ sites[15, 25]. The neutron powder diffraction analysis indicates that except Zn atoms are always accommodated in the 12$d$ sites, while site preference of both Co and Mn atoms are affected by the Mn content[26]. The occupation disorder of the Co, Zn and Mn atoms in the compound will affect the magnetic and transport properties, making the varying Mn content samples attractive to study. To understand the magnetic properties of obtained single crystals for different Mn contents, temperature dependence of the magnetic susceptibility $\chi$ (M-T curve) of Co$_{9.24}$Zn$_{9.25}$Mn$_{1.51}$ and Co$_{9.02}$Zn$_{9.18}$Mn$_{1.80}$ single crystals were measured and shown in Fig.2(a)-(d), respectively. It can be seen that $\chi$ exhibits typical ferromagnetic behavior with Curie temperature $T_C$ around 400 K for Co$_{9.24}$Zn$_{9.25}$Mn$_{1.51}$ and 368 K for Co$_{9.02}$Zn$_{9.18}$Mn$_{1.80}$, respectively. To help understand the ferromagnetism in the system, we have done the modified Curies-Weiss fitting via the formula[27] $\frac{H}{M} = \frac{T}{C} - \frac{T_C}{C}$ as shown in the inset of Fig. 2(f), where $C$ is Curie-Weiss constant. The good fitting result yields $C$ = 20.358 and $T_C$ = 368 K for Co$_{9.02}$Zn$_{9.18}$Mn$_{1.80}$. It should be noted that the Curie temperature of the Co$_{9.24}$Zn$_{9.25}$Mn$_{1.51}$ sample is too high (440 K) to do the Curies-Weiss fitting due to the limited temperature range. Compared with the Curie transition temperature of Co$_{10}$Zn$_{10}$ ($T_C$ = 462 K)[15], the obtained $T_C$ in our present compounds are significantly reduced. This indicates that the doping of Mn atoms weakens the ferromagnetic exchange interaction in the crystals. We calculated the effective magnetic moment by using[27] $\mu_{eff} = 2.83\sqrt{C}$ and obtained 12.77 $\mu_B$/f.u. for Co$_{9.02}$Zn$_{9.18}$Mn$_{1.80}$. The obtained value here is much smaller than the reported $\mu_{eff}$ = 1.6 $\mu_B$ (equivalent to 19.2 $\mu_B$/f.u.) for Co$_8$Zn$_8$Mn$_4$ [28]. This indicates that the ratio of Co atoms and Mn atoms has a greater effect on the effective magnetic moment $\mu_{eff}$. In addition, upon cooling to around 120 K, the magnetic susceptibility $\chi$ decreases significantly with the decreasing temperature. This is because at low temperatures, the Co moment remains ordering, while the Mn moment remains fluctuating, resulting in disorder so causing eventually a gradual decrease in saturated magnetization[16].

Interestingly, there appears a sharp peak in both FCC and FCW curves around $T_C$ as shown in the insets of Fig. 2(a)-(c), and we defined the temperature point at which the peak occurs as $T_P$. However, this sharp peak disappears when applied 1 T field as shown in Fig. 2(e) and (f). This peak-like character, which appears in low fields but suppressed in high fields, was also observed in MnSi, $Fe_{5-x}GeTe_2$, $Cu_2OSeO_3$ and other skyrmion materials[2, 29-33]. This feature is generally considered as a significant indicator of the precursor for skyrmion phase[2]. To confirm the existence of skyrmions in the sample, we measured the magnetic field dependence of isothermal magnetization curves ($M$ vs $\mu_0H$). The differential of magnetization d$M$/d$H$ vs $\mu_0H$ near $T_P$ (360 K) for $Co_{9.02}Zn_{9.18}Mn_{1.80}$ sample displays a typical peak in the temperature range from 354 K to 360 K (Fig. 3(a)). This feature is a hallmark associated with the formation and annihilation of skyrmions with the evolution of applied magnetic field[15, 23]. Based on the sensitivity of ac susceptibility $x'$ for magnetic properties, we further measured $x'$ as a function of magnetic field for $Co_{9.02}Zn_{9.18}Mn_{1.80}$ at various temperatures around $T_P$ from 346 K to 361 K. As shown in Fig. 3(b), it can be seen that $x'$ increases rapidly with increasing magnetic field near $H_1$, at which the transition from helical to conical phase occurs[34]. In addition, the dip anomalies in the $x'$-$\mu_0H$ curves appear between $H_2$ and $H_3$ from 354 K to 360 K. In fact, this anomaly that looks like a small pocket proves the formation of skyrmion phase[11, 15, 23]. When the temperature is lower than 352 K, the first anomaly which represents the helical-to-conical transition still exists, but the pocket between $H_2$ and $H_3$ related to the skyrmion phase almost disappear. Meanwhile, all three anomalies disappear when the temperature is above 361 K. Therefore, the temperature ranges of the abnormality appearing in the $x'$-$\mu_0H$ curves and d$M$/d$H$ vs $\mu_0H$ curves are basically the same. In the inset of Fig. 3(b), we plot the phase diagram from the results of the ac susceptibility, clearly showing the evolution of the magnetic structure with applied magnetic field. With increasing magnetic field, the magnetic structure of the sample undergoes three transitions: from helical to conical phase, then enters the skyrmion phase, and finally reenters the conical phase. We demonstrate that skyrmions appear in a narrow temperature interval of about 6 K near $T_P$, which further indicates the high quality of our sample.

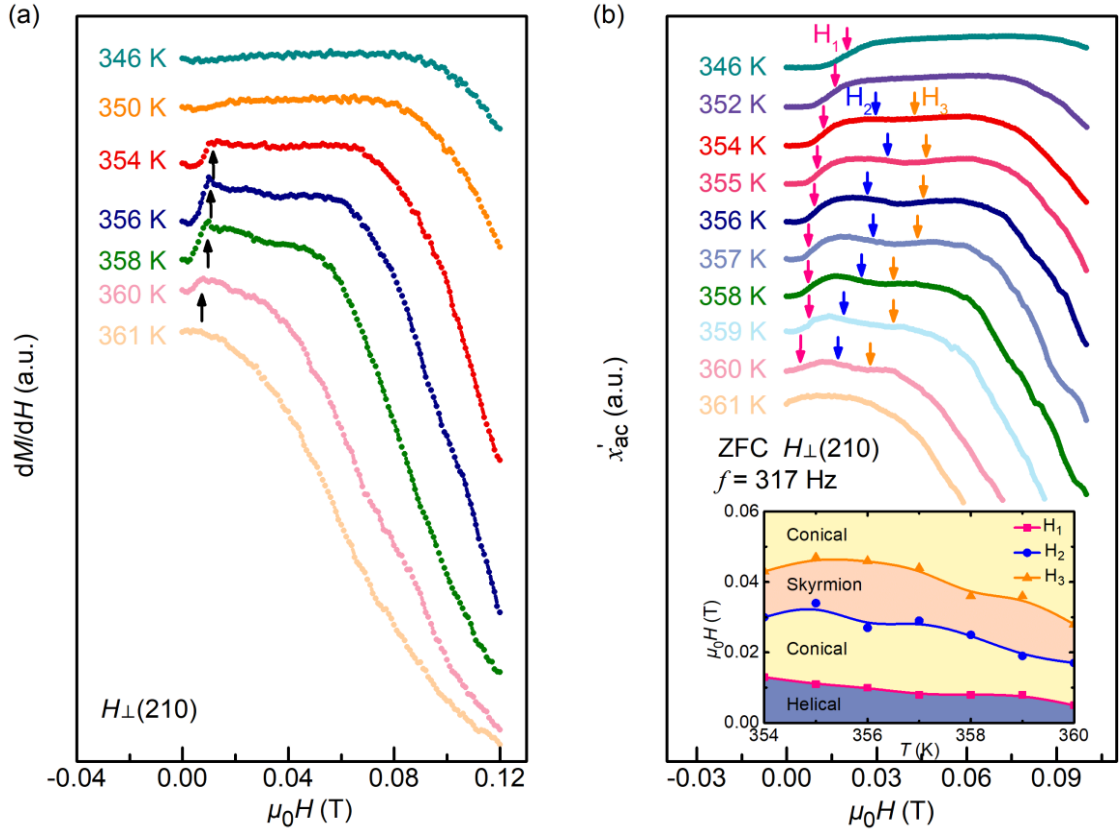

**Fig. 3.** (a) Isothermal $dM/dH$ versus $\mu_0 H$ curve for $Co_{9.02}Zn_{9.18}Mn_{1.80}$ sample. (b) Isothermal ac susceptibility as a function of field measured under excitation field of 1 Oe and 317 Hz ac magnetic field at various temperatures after ZFC process for $Co_{9.02}Zn_{9.18}Mn_{1.80}$ sample. Inset: Phase diagram of magnetic structure.

Fig. 4(a) and (b) show the temperature dependences of the zero-magnetic field longitudinal resistivity $\rho(T)$ of the $Co_{9.24}Zn_{9.25}Mn_{1.51}$ and $Co_{9.02}Zn_{9.18}Mn_{1.80}$ samples, respectively. It can be seen that both samples exhibit metallic behavior during the whole temperature range of 2 K ~ 380 K. However, the anomaly near 360 K is seen in the $\rho(T)$ curve of $Co_{9.02}Zn_{9.18}Mn_{1.80}$ compound at the almost same temperature point corresponding to the ferromagnetic transition in the *M(T)* curve as shown in Fig. 4(b). It is even particularly clear in the d$\rho$/d*T* curve. While the Curie temperature of the $Co_{9.24}Zn_{9.25}Mn_{1.51}$ compound is so high that it is out of present measurement range, no corresponding kink can be detected. Our finding of the kink in both *ρ(T)* and *M(T)* curves further confirms the high quality of our samples. Subsequently, we focused on the data of the resistivity in the low temperature region as shown in the insets of Fig. 4(a) and (b). The curve behavior indicates that the data at low temperature can be described by the Fermi liquid form[35]: $\rho = \rho_0 + AT^2$. Where $\rho_0$ is the residual resistivity and *A* is a measure of the strength of electron-electron scattering. The fit of our data to this form results in $\rho_0 = 174\ \mu\Omega\ cm$, and $A = 2.76 \times 10^{-3}\ \mu\Omega\ cm/K^2$ for $Co_{9.24}Zn_{9.25}Mn_{1.51}$ compound. And $\rho_0 = 201\ \mu\Omega\ cm$, and $A = 4.20 \times 10^{-3}\ \mu\Omega\ cm/K^2$ for $Co_{9.02}Zn_{9.18}Mn_{1.80}$ compound. The quadratic temperature dependence of *ρ* indicates that electron-electron scattering is the dominant temperature-dependent scattering mechanism for electrical transport below 20 K[35]. As the temperature increases, the *ρ(T)* gradually deviates from the fitted curve. This is because the disorder caused by the fluctuation of Mn moment affects scattering process of the conduction electrons, which eventually leads to the emergence of non-Fermi liquid (NFL) behavior[16, 36]. This result observed in the *ρ(T)* here is consistent with that observed gradual decrease of magnetization in the *M(T)* curves below 100 K (Fig.2(a)-(d)). Both are originating from the Mn disorder.

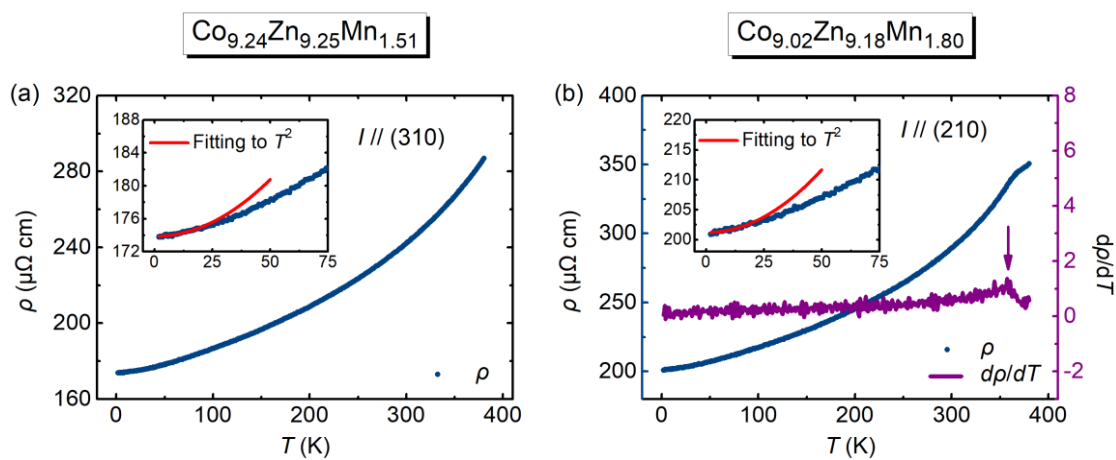

**Fig. 4.** The temperature dependence of the longitudinal resistivity ρ (blue line) without applied field. The purple line shows temperature-dependent differential resistivity of $Co_{9.02}Zn_{9.18}Mn_{1.80}$ sample. Insets show the results of fitting to $T^2$.

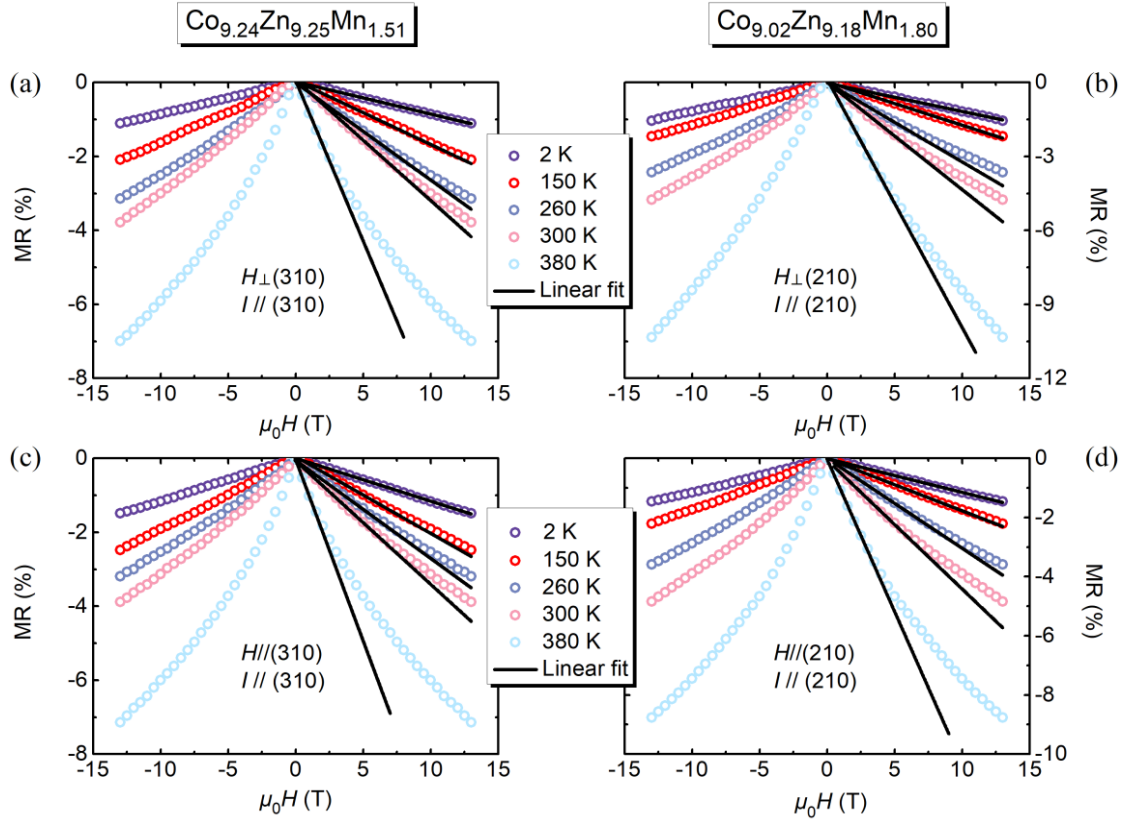

**Fig. 5.** Magnetic-field dependence of magnetoresistance at various temperatures. The black lines are the linear fits of MR.

Combined with the $M(T)$ and $\rho(T)$ measurement results, we measured the magnetoresistance (MR) of the $Co_{9.24}Zn_{9.25}Mn_{1.51}$ and $Co_{9.02}Zn_{9.18}Mn_{1.80}$ samples and the results are shown in Fig. 5. It can be seen that both the LMR and TMR displays negative over a wide range temperature from 2 K ~ 380 K. Moreover, the LMR and TMR of the samples show a perfect linear effect between 0 T ~ 13 T at $T = 2$ K. As the temperature increases, MR still maintains a linear behavior in the low-field region, but gradually deviates from linearity in the high-field region. Based on the current research progress on the MR effect, there exists two mechanisms for the negative MR behavior: (1) In heavy fermion compounds above the Kondo temperature scale, a regime where

*f*-electrons are not a part of the coherent excitations defined around a large Fermi surface and conduction electrons are thus scattered by local moments, the applied magnetic field would suppressed the electronic scattering rate from local moments and magnetic impurities, leading to negative simultaneous LMR and TMR[37]. (2) Simultaneous negative LMR and TMR are also observed in ferromagnetic compounds, where the applied magnetic field suppresses fluctuations of the ferromagnetic order parameter, thus decreases the inelastic scattering rate for conduction electrons[38]. Considering that $Co_{9.24}Zn_{9.25}Mn_{1.51}$ and $Co_{9.02}Zn_{9.18}Mn_{1.80}$ single crystals are metallic ferromagnet, leaving the suppressed fluctuations of ferromagnetic ordering as the mechanism behind[39, 40]. Furthermore, the linear effect of MR was generally speculated to the ferromagnetic correlation between the localized spins[41]. This is understandable for our present ferromagnetic $Co_{9.24}Zn_{9.25}Mn_{1.51}$ and $Co_{9.02}Zn_{9.18}Mn_{1.80}$ compounds.

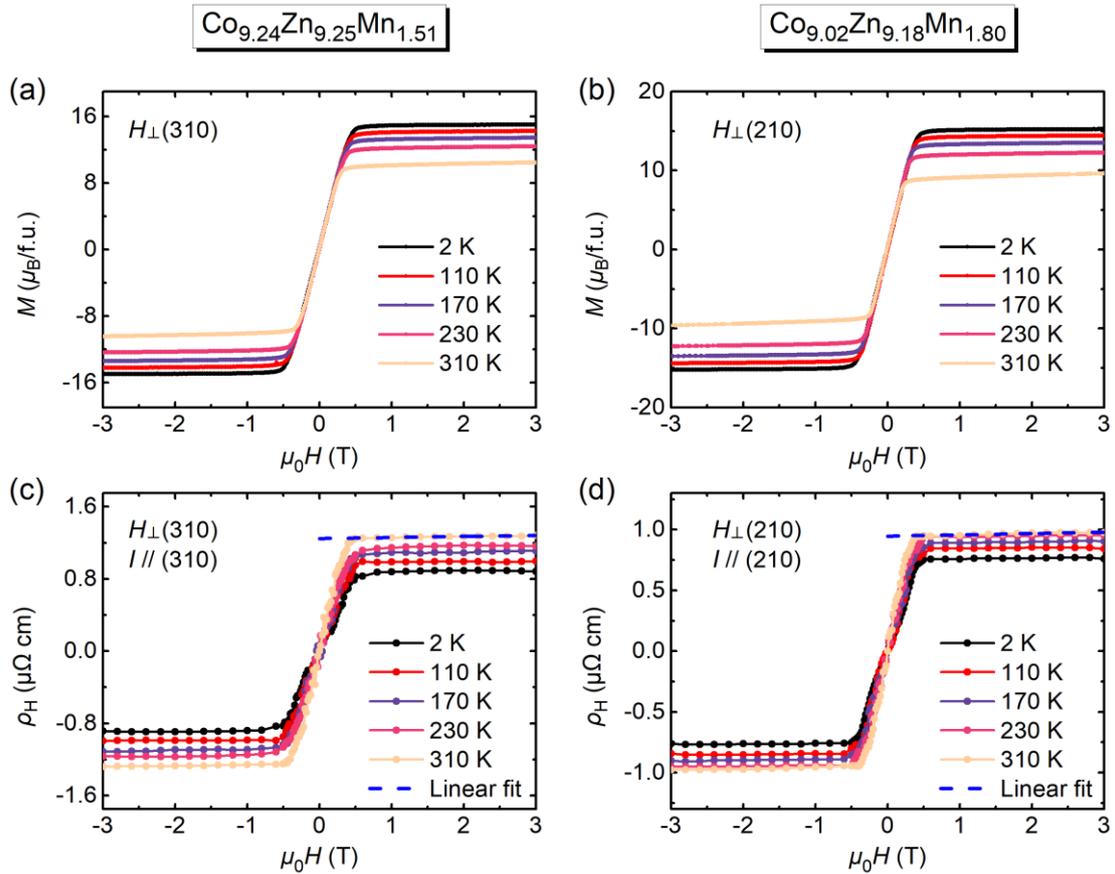

**Fig. 6.** (a) and (b) Magnetization $M$ as a function of applied field $\mu_0 H$ at various temperatures. (c) and (d) The Hall resistivity $\rho_H$ as a function of magnetic field $\mu_0 H$ for samples. The blue dash lines show the linear fit of $\rho_H$ versus $\mu_0 H$.

The transport properties and magnetic properties are closely correlated. The magnetization curves $M$-$\mu_0 H$ displays typical ferromagnetic behavior for both $Co_{9.24}Zn_{9.25}Mn_{1.51}$ and $Co_{9.02}Zn_{9.18}Mn_{1.80}$ compounds as shown in Fig. 6(a) and (b). With increasing temperature, the saturated magnetization increases continuously, and reaches its maximum value of about $14.93\ \mu_B/\text{f.u.}$ for $Co_{9.24}Zn_{9.25}Mn_{1.51}$ and $15.14\ \mu_B/\text{f.u.}$ for $Co_{9.02}Zn_{9.18}Mn_{1.80}$ samples at $T = 2$ K. The obtained value of $15.14\ \mu_B/\text{f.u.}$ for $Co_{9.02}Zn_{9.18}Mn_{1.80}$ sample here is comparable to that reported $Co_9Zn_9Mn_2$[16]. As the Hall mark of ferromagnetism, the Hall resistivity $\rho_H$ increases quickly to certain saturated values at low field, and reaches to almost linear with further increasing magnetic field at the $\rho_H$ vs $\mu_0 H$ curves under various temperatures (Fig.6(c) and (d)).

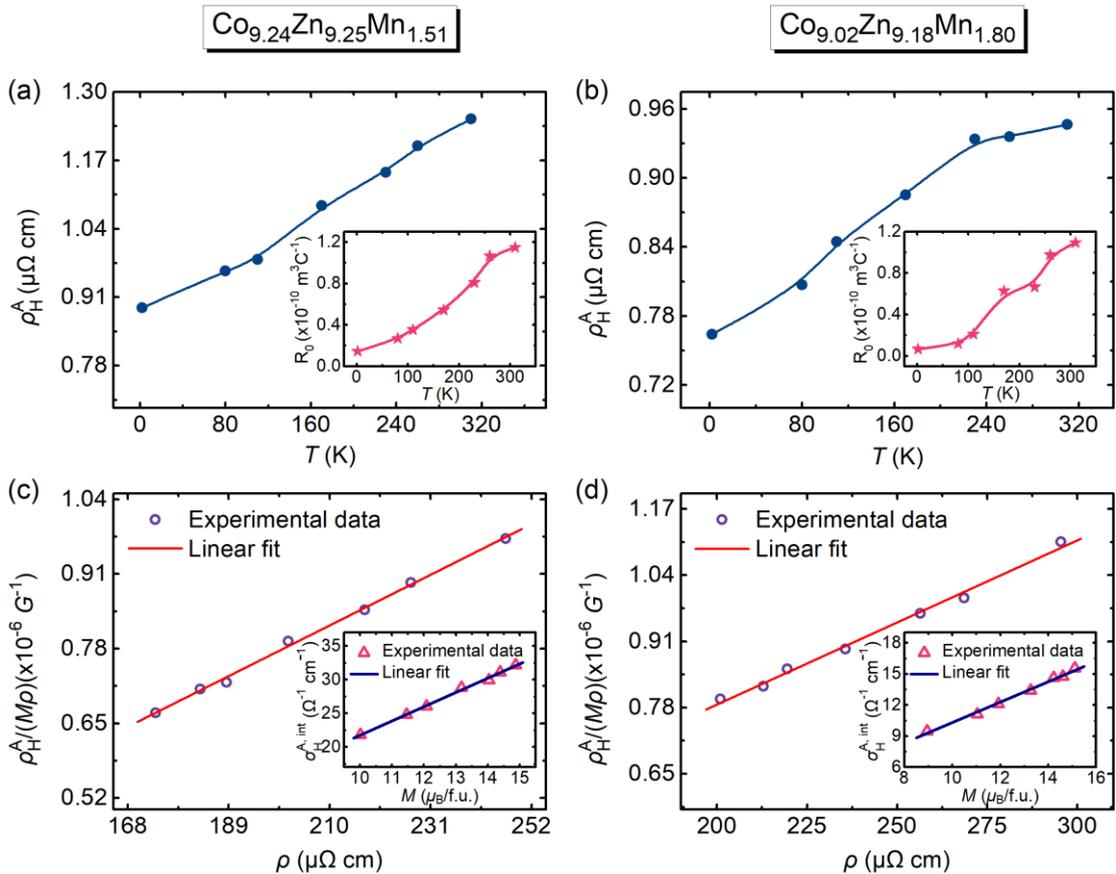

**Fig. 7.** (a) and (b) The temperature dependence of $\rho_H^A$ for samples. The insets show the temperature dependence of $R_0$. (c) and (d) Plot of $\rho_H^A/(M\rho)$ vs $\rho$. The insets show the intrinsic AHC as a function of $M$ for samples. The red solid lines and blue solid lines are the linear fits of the experimental data.

For ferromagnetic materials[24], Hall resistivity $\rho_H$ is the sum of the normal Hall resistivity $\rho_H^O$ and anomalous Hall resistivity $\rho_H^A$: $\rho_H = \rho_H^O + \rho_H^A = R_0 B + R_s \mu_0 M$. Here $R_0$ and $R_s$ are normal and anomalous Hall coefficient, respectively. The $\rho_H^A$ is typically obtained as the $H = 0$ intercept of a linear fit to $\rho_H$ versus $\mu_0 H$ at high field region, and the slope correspond to $R_0$. The linear fitting is shown in Fig. 6(c) and (d), and the results are shown in Fig. 7(a) and (b). It can be seen that the $\rho_H^A$ values increase gradually with increasing temperature. $R_0$ are positive, indicating the majority carriers are hole (Fig. 7(a) and (b)). The current view believes that there are three mechanisms of AHE, namely: skew scattering, side-jump scattering and intrinsic mechanism[42-47]. The skew scattering and side-jump scattering are also called extrinsic mechanism, which are generally considered to be the result of asymmetric scattering of conducting electrons affected by spin-orbit interaction[43-45]. While, the intrinsic mechanism, also known as K-L theory, is related to the Berry curvature of electronic band structure in momentum space[46, 47]. According to previous reports, the dominant mechanism can be decided by the formula[48, 49] $\rho_H^A = a(M)\rho + b(M)\rho^2$. The first term corresponds to the skew scattering contribution, and the second term represents the intrinsic mechanism or side-jump scattering contribution[49]. The skew scattering contribution a(*M*) is usually linear with magnetization[50]. Accordingly, a(*M*) can be obtained by plotting $\rho_H^A/(M\rho)$ vs $\rho$, i.e. the skew scattering contribution can be obtained. As previously mentioned, the skew scattering generally only dominates the AHE in the high-purity regime[24], while the order of magnitude of $\sigma_{xx}$ of our samples is $10^3 (\Omega\text{ cm})^{-1}$, which is far less than $10^6 (\Omega\text{ cm})^{-1}$. Therefore, our samples do not belong to the high-purity regime, the skew scattering was ruled out as the main mechanism of AHE. Furthermore, we need to analyze the data of anomalous Hall conductivity after subtracting the skew-scattering contribution. Interestingly, the plotting $\rho_H^A/(M\rho)$ vs $\rho$ displays a wonderful linear behavior as shown in Fig. 7(c) and (d), which indicates that the intrinsic mechanism is the dominant mechanism of AHE in our present compounds[49]. To further confirm the intrinsic mechanism, we subtract the skew-scattering contribution and found that the obtained intrinsic AHC $\sigma_H^{A,int}$ (= $\frac{\rho_H^{A,int}}{\rho^2}$, $\rho_H^{A,int}$ is the intrinsic anomalous Hall resistivity) exhibits the linear dependence of $M$ as shown in the insets of Fig. 7(c) and (d). This is in good agreement with the K-L theory, which further confirms that the intrinsic mechanism plays a dominant role of the AHE in our samples[42, 48, 49]. The determination of the mechanism of AHE will help us to explore the possibility of $Co_9Zn_9Mn_2$ samples in novel electronic devices based on AHE.

## IV. CONCLUSION

In conclusion, we have successfully synthesized the chiral $\beta$-Mn-type $Co_{9.24}Zn_{9.25}Mn_{1.51}$ and $Co_{9.02}Zn_{9.18}Mn_{1.80}$ single crystals and investigated systematically magnetic and transport properties. A typical characteristic of skyrmion phase in a narrow temperature interval of about 6 K was observed above the room temperature via dc and ac magnetic susceptibility measurements. A negative linear magnetoresistance over a wide temperature range from 2 K to 380 K is observed and attributed to the suppression of the magnetic ordering fluctuation under high fields. The normal Hall coefficient $R_0$ is positive, indicating the hole carriers in the compounds. In addition, through quantitative analysis of AHE, we found that the intrinsic mechanism dominates the AHE in present lightly Mn-doped $Co_9Zn_9Mn_2$ instead of the previously reported skew scattering. Our study provides a basic understanding of the fundamental properties of $\beta$-Mn-type $Co_xZn_yMn_z$ that host the room-temperature skyrmions, which are useful for applications of skyrmion-based spintronic.

## ACKNOWLEDGMENTS

This work is jointly supported by the Ministry of Science and Technology of the People's Republic of China (2018YFB0704402). This work is supported by Key Research Project of Zhejiang Lab (No. 2021PE0AC02).